\documentclass[preprint,showpacs,preprintnumbers,amsmath,amssymb,nofootinbib,showkeys,aps]{revtex4}
\pdfoutput=1
\usepackage{epsfig}

\usepackage{graphicx}
\usepackage{dcolumn}
\usepackage{bm}
\usepackage{amsmath}
\usepackage{latexsym}
\usepackage{color}
\usepackage{hyperref}
\usepackage{mathrsfs}
\DeclareMathOperator{\atan}{arctan}

\newcommand{\be}{\begin{equation}}
\newcommand{\ee}{\end{equation}}
\newcommand{\texpdf}{\texorpdfstring}
\newcommand{\like}{\mathscr{L}}

\begin{document}


\title{Model independent constraints on transition redshift}

\author{J. F. Jesus$^{1,2}$}\email{jfjesus@itapeva.unesp.br}
\author{R. F. L. Holanda$^{3,4,5}$} \email{holanda@uepb.edu.br}
\author{S. H. Pereira$^{2}$} \email{shpereira@feg.unesp.br}

\affiliation{$^1$Universidade Estadual Paulista (Unesp), Campus Experimental de Itapeva - R. Geraldo {Alckmin 519}, 18409-010, Itapeva, SP, Brazil,
\\$^2$Universidade Estadual Paulista (Unesp), Faculdade de Engenharia, Guaratinguet\'a, Departamento de F\'isica e Qu\'imica - Av. Dr. Ariberto Pereira da Cunha 333, 12516-410, Guaratinguet\'a, SP, Brazil
\\$^3$Departamento de F\'isica Te\'orica e Experimental, Universidade Federal do Rio Grande do Norte, 59078-970, Natal - RN, Brazil,
\\$^4$Departamento de F\'{\i}sica, Universidade Federal de Sergipe, 49100-000, S\~ao Cristov\~ao - SE, Brazil,
\\$^5$Departamento de F\'{\i}sica, Universidade Federal de Campina Grande, 58429-900, Campina Grande - PB, Brazil}


\def\zt{\mbox{$z_t$}}

\begin{abstract}
This paper aims to put constraints on the transition redshift $z_t$, which determines the onset of cosmic acceleration, in cosmological-model independent frameworks. In order to perform our analyses, we consider a flat universe and {assume} a parametrization for the comoving distance $D_C(z)$ up to third degree on $z$, a second degree parametrization for the Hubble parameter $H(z)$ and a linear parametrization for the deceleration parameter $q(z)$. For each case, we show that {type Ia supernovae} and $H(z)$ data complement each other on the parameter {space} and tighter constrains for the transition redshift are obtained. By {combining} the type Ia supernovae  observations and Hubble parameter measurements it is possible to constrain the values of $z_t$, for each approach, as $0.806\pm 0.094$, $0.870\pm 0.063$ and $0.973\pm 0.058$ at 1$\sigma$ c.l., respectively. Then, such approaches provide cosmological-model independent estimates for this  parameter.

\end{abstract}

\maketitle



\section{Introduction} 

The idea of a late time accelerating universe is indicated by type Ia supernovae (SNe Ia) observations \cite{SN1,SN2,SN3,SN4,SN5,union,union2,union21} and confirmed by other independent observations such as Cosmic Microwave Background (CMB) radiation \cite{WMAP1,WMAP2,planck}, Baryonic Acoustic Oscillations (BAO) \cite{BAO1,BAO2,BAO3,BAO4,BAO5} and Hubble parameter, $H(z)$, measurements \cite{Omer,Omer2,Omer3}. The simplest theoretical model supporting such accelerating phase is based on a cosmological constant $\Lambda$ term \cite{CC,padmana} plus a Cold Dark Matter component \cite{reviewDM,bookDM,weinberg2013}, the so-called $\Lambda$CDM model. The cosmological parameters of such model have been constrained more and more accurately \cite{planck,Omer2,sharov} as new observations are added. Beyond a constant $\Lambda$ based model, several other models have been also suggested recently in order to explain the accelerated expansion. The most popular ones are based on a dark energy fluid \cite{peebles,sahni2} endowed with a negative pressure filling the whole universe. The nature of such exotic fluid is unknown, sometimes attributed to a single scalar field 
or even to mass dimension one fermionic fields. 
There are also modified gravity theories that correctly {describe} an accelerated expansion of the {Universe}, such as: massive gravity {theories} \cite{volkov}, 
modifications of Newtonian theory {(MOND)} \cite{mond,mond2},
 $f(R)$ and $f(T)$ theories that generalize the general {relativity} \cite{fR,fR3,fR4},
 models based on extra dimensions, as brane world {models} \cite{randall,pomarol,langlois,shir,cline}, 
string \cite{string} 
and Kaluza-Klein {theories} \cite{klein}, 
among others. Having adopted a particular model, the cosmological parameters can be determined {by using} statistical analysis of observational data. 

However, some works have tried to explore the history of the universe without to appeal to any specific cosmological model. Such approaches are sometimes called cosmography or cosmokinetic models \cite{kine1,kine2,kine3,kine4,kine5,kine6}, and we will refer to them simply as kinematic models. {This nomenclature} comes from the fact that the complete study of the expansion of the {Universe} (or its kinematics) is described just by the Hubble expansion rate $H = \dot{a}/a$, the deceleration parameter $q=-a\ddot{a}/\dot{a}^2$ and the jerk parameter $j = -\dddot{a}a^3/(a\dot{a}^3)$, where $a$ is the scale factor in the Friedmann-Roberson-Walker (FRW) metric. The only assumption is that space-time is homogeneous and isotropic. In such parametrization, a simple dark matter dominated universe has $q=1/2$ while the accelerating $\Lambda$CDM model has $j=-1$. The deceleration parameter allows to study the transition from a decelerated phase to an accelerated one, while the jerk parameter allows to study departures from the cosmic concordance model, without restricting to a specific model.

{Concerning} the deceleration parameter, {several studies have attempted to {estimate} at which redshift $z_t$ the universe undergoes a transition to accelerated phase \cite{Omer2,Omer3,kine3,cunha2008,guimaraes2009,Rani2015,limajesus2014,moresco2016}}. A model independent determination of the present deceleration parameter $q_0$ and deceleration-acceleration transition redshift $z_t$ is of fundamental importance in modern cosmology. As a new cosmic {parameter \cite{limajesus2014}}, it should be used to test several cosmological models. 

{In order to study the deceleration parameter in a cosmological-model independent framework}, it is {necessary} to use some parametrization for it. This methodology has both advantages and
disadvantages. One advantage is that it is independent of the matter and energy
content of the universe.  One disadvantage of this formulation is that it does not explain the cause of
the accelerated expansion. Furthermore, the value of the present deceleration parameter may depend on the assumed form of $q(z)$. One of the first analyses and constraints on cosmological parameters of kinematic models was done by {Elgar\o{}y and Multam\"aki} \cite{kine5} by employing Bayesian marginal likelihood analysis. Since {then,} several authors have implemented the analysis by including new data sets and also different parametrizations for $q(z)$.

For a linear parametrization, $q(z)=q_0+q_1 z$, the values for $q_0$ and $z_t$ found by Cunha and Lima \cite{cunha2008} were $q_0 \sim -0.7$, $z_t=0.43^{+0.09}_{-0.05}$ from 182 SNe Ia of {Riess {\it et al.}} \cite{SN4}, $-1.17 \leq q_0 \leq −0.16$, $z_t = 0.61^{+3.68}_{-0.21}$ from SNLS data set \cite{SN3} and $-1.0 \leq q_0 \leq −0.36$, $z_t = 0.60^{+0.28}_{-0.11}$ from Davis {{\it et al}.} data set \cite{SN5}. Guimar\~aes, Cunha and Lima \cite{guimaraes2009}  found $q_0=-0.71 \pm 0.21$ and $z_t=0.49^{+0.27}_{-0.09}$ using a sample of 307 SNe Ia from Union compilation \cite{union}. Also, for the linear parametrization, Rani {{\it et al}}. \cite{Rani2015} found $q_0 = -0.52 \pm 0.12$ and $z_t \approx 0.98$ using a joint analysis of age of galaxies, strong gravitational lensing
and SNe Ia data. For a parametrization of type $q(z)=q_0+q_1 z/(1+z)$,
Xu, Li and Lu \cite{Xu2009} used 307 SNe Ia together {with BAO and $H(z)$} data and found $q_0=-0.715 \pm 0.045$ and $z_t=0.609^{+0.110}_{-0.070}$. For the same parametrization, Holanda, Alcaniz and Carvalho \cite{holanda2013} found $q_0 = −0.85^{+1.35}_{-1.25}$ by using galaxy clusters of elliptical morphology based on their Sunyaev-Zeldovich effect (SZE) and X-ray observations. Such parametrization has also been studied in \cite{kine5,cunha2008}.

{As one may see, the determination of the deceleration parameter is a relevant subject in modern cosmology, as well as the determination of the Hubble parameter $H_0$.} In seventies, Sandage \cite{sandage} foretold that the determination of $H_0$ and $q_0$ would be the main role of cosmology for the forthcoming decades. The inclusion of the transition redshift $z_t$ as a new cosmic discriminator has been advocated by some authors \cite{limajesus2014}. An alternative method to access the cosmological parameters in a model independent fashion is by means of the study of {$H(z)=\dot{a}/a=-[1/(1 +
z)] dz/dt$}. In the so called \emph{cosmic chronometer} approach,
the quantity $dz$ is obtained from spectroscopic surveys and
the only quantity to be measured is the differential age evolution of the universe ($dt$) in a given
redshift interval ($dz$). By using the results from Baryon Oscillation Spectroscopic Survey {(BOSS) \cite{BAO3,BAO4,BAO5}, Moresco {\it et al}.} \cite{moresco2016} have obtained a cosmological-model independent determination of the transition redshift as $z_t = 0.4 \pm 0.1$ (see also \cite{Omer,Omer2,Omer3}).

In the present work we study the transition redshift by means of a third order parametrization of the {comoving} distance, a second order parametrization of $H(z)$ and a linear parametrization of $q(z)$. By combining luminosity distances from SNe Ia \cite{JLA} and $H(z)$ measurements, it is possible to determine $z_t$ values in these cosmological-model independent frameworks. {In such approach we obtain an interesting complementarity between the observational data and, consequently, tighter constraints on the parameter spaces.}

The paper is organized as follows. In Section {\ref{basic},} we present the basic equations concerning the obtainment of $z_t$ from luminosity distance, $H(z)$ and $q(z)$. Section {\ref{samples} presents} the data set used and the analyses are presented in Section \ref{analysis}. Conclusions are left to Section \ref{conclusion}.

\section{\label{basic}Basic equations}

Let us discuss (from a more observational viewpoint) the possibility to enlarge Sandage's vision by including the
transition redshift, $z_t$, as the third cosmological number.
To begin with, consider the general expression for the deceleration parameter $q(z)$ as given by: 
\begin{equation}
 q(z)=-\frac{\ddot{a}}{aH^2}=\frac{1+z}{H}\frac{dH}{dz}-1\,, \label{qzH}
\end{equation}
from which the transition redshift, $z_t$, can be defined as $q(z_t)=0$, leading to:
\begin{equation}
z_t = \left[\frac{d {\ln} H(z)}{dz}\right]^{-1}_{|_{z=z_t}} - 1.
\label{z_tM}
\end{equation}

Let us assume a flat Friedmann-Robertson-Walker cosmology. In such a framework, the luminosity distance, $d_L$ (in Mpc), is given by:
\begin{equation}
\label{dl}
d_L(z)=(1+z)d_C(z),
\end{equation}
where $d_C$ is the comoving distance:
{
\begin{equation}
\label{dc}
d_C(z) = c\int_0^z \frac{dz'}{H(z')},
\end{equation}
with} $c$ being the speed of light in km/s and $H(z)$ the Hubble parameter in km/s/Mpc. For mathematical convenience, we choose to work with dimensionless quantities. Then, we define the dimensionless distances, $D_C\equiv\frac{d_C}{d_H}$, $D_L\equiv\frac{d_L}{d_H}$, $d_H\equiv c/H_0$ and the dimensionless Hubble parameter, $E(z)\equiv\frac{H(z)}{H_0}$. Thus, we have:
\begin{equation}
D_L(z)=(1+z)D_C(z),
\label{Dl}
\end{equation}
and
{
\begin{equation}
\label{Dc}
D_C(z) = \int_0^z \frac{dz'}{E(z')},
\end{equation}
from which follows
\begin{equation}
 E(z)=\bigg[\frac{dD_C(z)}{dz}\bigg]^{-1},
 \label{EzDc}
\end{equation}
}
From \eqref{z_tM} we also have:
\begin{equation}
\label{ztfromHz}
z_t =\left[\frac{d\ln E(z)}{dz}\right]_{|z=z_t}^{-1} -1.
\end{equation}

{Therefore, from a formal point of view, we may access the value of $z_t$ through  a parametrization of both $q(z)$ and $H(z)$, at least around a redshift interval involving the transition redshift.} As a third method we can also parametrize the co-moving distance, which is directly related to the luminosity distance, in order to study the transition redshift. In which follows we present the three different methods considered here.

\subsection{\texpdf{$z_t$}{zt} from comoving distance, \texpdf{$D_C(z)$}{Dc(z)}}

In order to put limits on $z_t$ by considering the comoving distance, we can write $D_C(z)$ by a third degree polynomial such as: 
\begin{equation}
 D_C = z + d_2 z^2 + d_3 z^3.
 \label{DcPolin}
\end{equation}
where $d_2$ and $d_3$ are free parameters. Naturally, from Eqs. (\ref{EzDc}) and (\ref{DcPolin}), one obtains
\begin{equation}
E(z)=\frac{1}{1 + 2d_2 z + 3d_3 z^2}.
\label{Ezdcpolin}
\end{equation}
Solving Eq. (\ref{ztfromHz}) with $E(z)$ given by (\ref{Ezdcpolin}), we find
\begin{equation}
 z_t=\frac{-2d_2-3d_3\pm\sqrt{4d_2^2-6d_2d_3+9d_3^2-9d_3}}{9d_3},
\end{equation}
where we may see there are two possible solutions to $z_t$. From a statistical point of view, aiming to constrain $z_t$, maybe it is better to write the coefficient $d_3$ in terms of $z_t$ and $d_2$ via Eq. (\ref{ztfromHz}) as
\begin{equation}
\label{d3}
d_3(d_2,z_t)=-\frac{1 + 2 d_2 + 4 d_2 z_t}{3 z_t (2 + 3 z_t)}.
\end{equation}
Then
\begin{equation}
\label{dc5}
E(z)=\left[1 + 2 d_2 z - \frac{1 + 2 d_2 + 4 d_2 z_t}{z_t (2 + 3 z_t)} z^2\right]^{-1}.
\end{equation}
Finally, from Eqs. (\ref{Dl}), (\ref{DcPolin}) and (\ref{d3}) the dimensionless luminosity distance is
\begin{equation}
\label{dl2}
D_L(z)= (1+z)\left[z + d_2z^2 - \frac{1 + 2 d_2 + 4 d_2 z_t}{3 z_t (2 + 3 z_t)} z^3\right].
\end{equation}
Equations (\ref{dl2}) and (\ref{dc5}) are to be compared with luminosity distances from SNe Ia and $H(z)$ measurements, respectively, in order to determine $z_t$ and $d_2$. 

\subsection{\texpdf{$z_t$}{zt} from \texpdf{$H(z)$}{Hz}}
 
In order to assess $z_t$ from Eq. (\ref{z_tM}) by {means} of $H(z)$ we need an expression for $H(z)$. If one wants to avoid dynamical assumptions, one must to resort to kinematical methods which uses an expansion of $H(z)$ over the redshift.

The simplest expansion of $H(z)$ over the {redshift, the linear expansion, gives no transition. To realize this, let us take}
\begin{equation}
\frac{H(z)}{H_0}=E(z)=1+h_1z .
\end{equation}
From (\ref{ztfromHz}), we have
\begin{equation}
 1+z_t=\left[\frac{d\ln E(z)}{dz}\right]^{-1}_{z=z_t}=\frac{1}{h_1}+z_t\Rightarrow\frac{1}{h_1}=1.
 \label{1z_tM}
\end{equation}
Therefore, the transition redshift is undefined in this case. 

Let us now try the next simplest $H(z)$ expansion, namely, the quadratic expansion:
\begin{equation}
 \frac{H(z)}{H_0} =E(z)= 1 + h_1z + h_2z^2.
 \label{EzPolin}
\end{equation}
In this case, inserting \eqref{Ezdcpolin} into (\ref{1z_tM}), we are left with:
\begin{equation}
 1+z_t=\frac{1 + h_1z_t + h_2z_t^2}{h_1 + 2h_2z_t}\Rightarrow(1+z_t)(h_1 + 2h_2z_t)=1+h_1z_t+h_2z_t^2,
\end{equation}
from which follows an equation for $z_t$:
\begin{equation}
 h_2z_t^2+2h_2z_t+h_1-1=0,\label{eq19}
\end{equation}
whose solution is:
\begin{equation}
 z_t=-1 \pm\sqrt{1+\frac{1-h_1}{h_2}}.
\end{equation}
We may exclude the negative root, which would give $z_t<-1$ and this value is not possible (negative scale factor). Thus, if one obtains the $h_1$ and $h_2$ coefficients from a fit to $H(z)$ data, one may obtain a model independent estimate of transition redshift from
\begin{equation}
 z_t=-1 + \sqrt{1+\frac{1-h_1}{h_2}}.
 \label{ztHz2}
\end{equation}

Equation (\ref{ztHz2}) already is an interesting result, and shows the reliability of the quadratic model as a kinematic assessment of transition redshift. {It is easy to see that taking $h_2=0$ into (\ref{eq19}) does not furnish any information about the transition redshift.}

In order to constrain the model with SNe Ia data, we obtain the luminosity distance from Eqs. (\ref{Dl}), (\ref{Dc}) and (\ref{EzPolin}). We have
{
\begin{equation}
 D_C=\int_0^z\frac{dz'}{E(z')}=\int_0^z\frac{dz'}{1 + h_1z' + h_2z'^2},
\end{equation}
}
which gives three possible solutions, according to the sign of $\Delta\equiv h_1^2-4h_2$ such as
\begin{eqnarray}
 D_C=\left\{\begin{array}{ll}
             \dfrac{2}{\sqrt{-\Delta}}\left[\atan\left(\dfrac{2h_2z+h_1}{\sqrt{-\Delta}}\right)-\atan\dfrac{h_1}{\sqrt{-\Delta}}\right],&\Delta<0\\
             \dfrac{2z}{h_1z+2},&\Delta=0\\
             \dfrac{1}{\sqrt{\Delta}}\ln\left|\left(\dfrac{\sqrt{\Delta}+h_1}{\sqrt{\Delta}-h_1}\right)\left(\dfrac{\sqrt{\Delta}-h_1-2h_2z}{\sqrt{\Delta}+h_1+2h_2z}\right)\right|,&\Delta>0
            \end{array}
 \right.\label{eqDCz}
\end{eqnarray}

However, in order to obtain the likelihood for the transition redshift, we must reparametrize the Eq. (\ref{EzPolin}) to show its explicit dependency on this parameter. Notice also that from Eq. (\ref{ztHz2}) we may eliminate the parameter $h_1$:
\begin{equation}
h_1 = 1 + h_2[1-(1+z_t)^2]\,,\label{eqh1}
\end{equation}
thus we may write $D_C(z)$ from \eqref{eqDCz} just in terms of $z_t$ and $h_2$, from which follows the luminosity distance $D_L=D_C(z)(1+z)$.

\begin{figure}[t]
\centering
 \includegraphics[width=.49\linewidth]{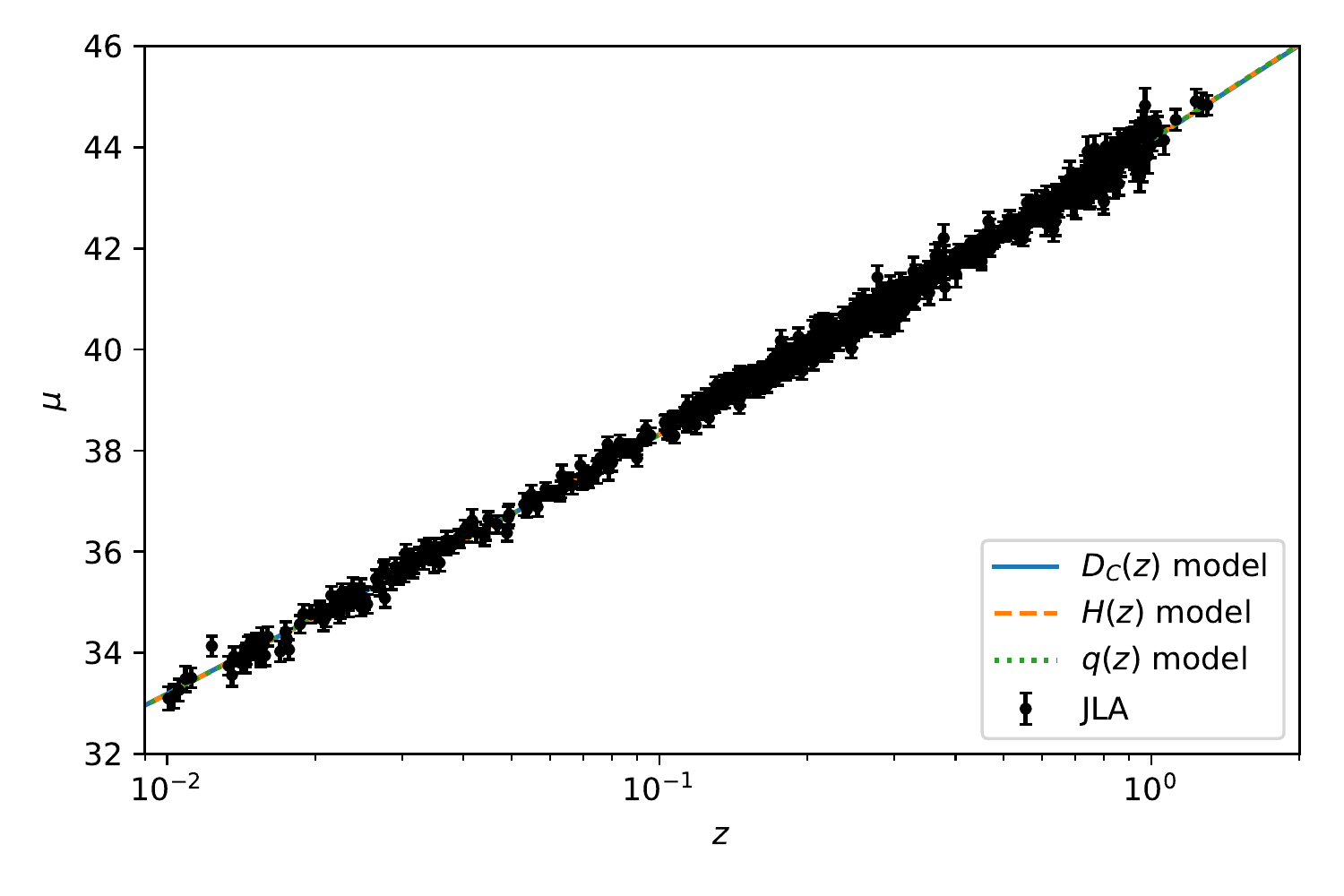}
 \includegraphics[width=.49\linewidth]{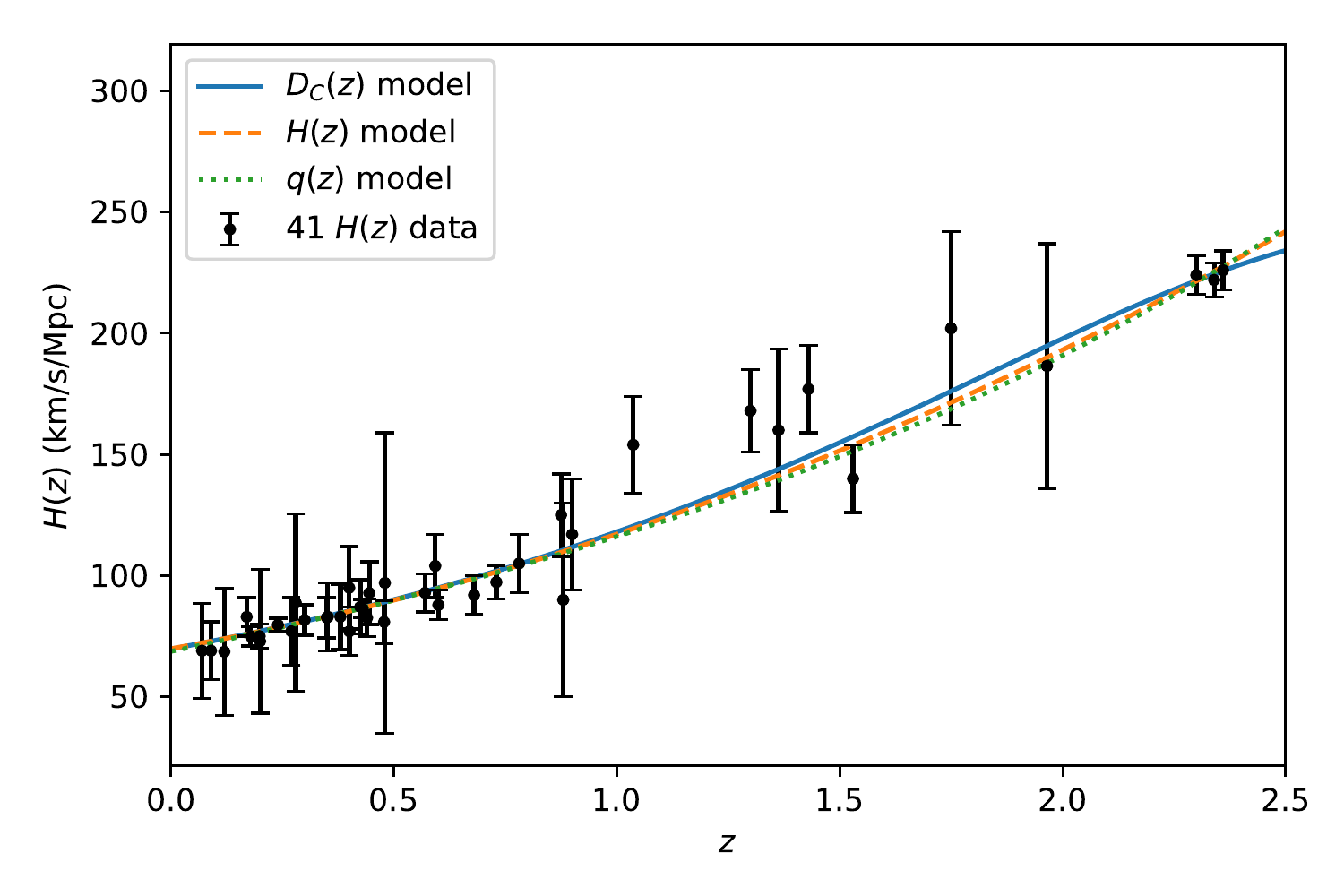}
 \caption{\label{Data} {\bf a)} SNe Ia distance moduli from JLA. The data for $\mu$ were estimated from Eq. (\ref{muBet}), with SNe Ia parameters from the $D_C(z)$ model (Table \ref{tab1}) and the error bars comes from the diagonal of the covariance matrix. The lines represent the best fit from SNe+$H(z)$ data for each model. {\bf b)} 41 $H(z)$ data compilation. The lines represent the best fit from SNe+$H(z)$ data for each model.}
 \label{AllData}
\end{figure}

\subsection{\texpdf{$z_t$}{zt} from \texpdf{$q(z)$}{qz}}

Now let us see how to assess $z_t$ by means of a parametrization of $q(z)$. From (\ref{qzH}) one may find $E(z)$
\begin{equation}
 E(z)=\exp\left[\int_0^z\frac{1+q(z')}{1+z'}dz'\right].
\end{equation}
If we assume a linear $z$ dependence in $q(z)$, as
\begin{equation}
 q(z)=q_0+q_1z,\label{eqqz}
\end{equation}
which is the simplest $q(z)$ parametrization that allows for a transition, one may find
\begin{equation}
 E(z)=e^{q_1z}(1+z)^{1+q_0-q_1},\label{eqEzq}
\end{equation}
while the {comoving} distance $D_C(z)$ (\ref{Dc}) is given by
\begin{equation}
 D_C(z)=e^{q_1}q_1^{q_0-q_1}\left[\Gamma(q_1-q_0,q_1)-\Gamma(q_1-q_0,q_1(1+z))\right],
\end{equation}
where $\Gamma(a,x)$ is the incomplete gamma function defined by \cite{AbraSteg} as $\Gamma(a,x)\equiv\int_x^\infty e^{-t}t^{a-1}dt$, with $a>0$.

From \eqref{eqqz} (or from \eqref{eqEzq} and \eqref{ztfromHz}) it is easy to find:
\begin{equation}
z_t=-\frac{q_0}{q_1} \hspace{1cm}\textrm{and}\hspace{1cm} q_0=-q_1 z_t,
\end{equation}
from which {follows $D_C(z)$ and} $D_L(z)$ as a function of just $z_t$ and $q_1$, which can be constrained from observational data.

\section{\label{samples}Samples}
\subsection{\texpdf{$H(z)$}{Hz} data}
{Hubble parameter data in terms of redshift yields one of the most straightforward cosmological tests because it is inferred from astrophysical observations alone, not depending on any background cosmological models.}

At the present time, the most important methods for obtaining $H(z)$ data are\footnote{See \cite{limajesus2014} for a review.} (i) through ``cosmic chronometers", for example, the differential age of galaxies (DAG), (ii) measurements of peaks of acoustic oscillations of baryons (BAO) and (iii) through correlation function of luminous red galaxies (LRG).

The data we work here are a combination of the compilations from Sharov and Vorontsova \cite{SharovVor14} and Moresco {\it et al.} \cite{moresco2016} as described on Jesus {\it et al.} \cite{JesusHz41}. Sharov and Vorontsova \cite{SharovVor14} added 6 $H(z)$ data in comparison to Farooq and Ratra \cite{Omer2} compilation, which had 28 measurements. Moresco {\it et al.} \cite{moresco2016}, on their turn, have added 7 new $H(z)$ measurements in comparison to Sharov and Vorontsova \cite{SharovVor14}. By combining both datasets, Jesus {\it et al.} \cite{JesusHz41} have arrived at 41 $H(z)$ data, as can be seen on Table 1 of \cite{JesusHz41} and {Figure \ref{AllData}b here}.

\subsection{JLA SNe Ia compilation}

The JLA compilation \cite{JLA} consists of 740 SNe Ia from the SDSS-II \cite{Sako2014} and SNLS \cite{C11}
collaborations. Actually, this compilation produced {recalibrated} SNe Ia light-curves and associated distances for the SDSS-II and SNLS samples in order to {improve} the accuracy of cosmological constraints, limited by systematic measurement uncertainties, as, for instance, the uncertainty in the band-to-band and survey-to-survey relative flux calibration. The  light curves have high quality and were obtained by using {an improved SALT2 (Spectral Adaptive Light-curve Templates) method \cite{JLA,guy2007,guy2010,betoule2014}.} The data set includes several low-redshift samples ($z < 0.1$), all three seasons from the SDSS-II ($0.05 \leq z \leq 0.4$) and three years from SNLS ($0.2 < z < 1.4$). {See Fig. \ref{AllData}a and more details in next section.}

\section{\label{analysis}Analyses and Results}
In our analyses, we have used flat priors over the parameters, so the posteriors are always proportional to the likelihoods. 
For $H(z)$ data, the likelihood distribution function is given by $\like_H \propto e^{-\frac{\chi^2_H}{2}}$, where 
\begin{equation}
\chi^2_H = \sum_{i = 1}^{41}\frac{{\left[ H_{obs,i} - H(z_i,H_0,z_t,\theta_{mod,j})\right] }^{2}}{\sigma^{2}_{H_i,obs}} ,
\label{chi2H}
\end{equation}
where $\theta_{mod,j}$ is the specific parameter for each model, {namely} $d_2$, $h_2$ or $q_1$, for $D_C(z)$, $H(z)$, $q(z)$ parametrizations, respectively. 

As explained on \cite{JLA}, we may assume that supernovae with identical color, shape and galactic environment have, on average, the same intrinsic luminosity for all redshifts. In this case, the distance modulus $\mu=5\log_{10}(d_L(\mathrm{pc})/10)$ may be given by
\begin{equation}
 \mu=m_B^*-(M_B-\alpha\times X_1+\beta\times C)
 \label{muBet}
\end{equation}
where $X_1$ describes the time stretching of the light-curve, $C$ describes the supernova color at maximum brightness, $m_B^*$ corresponds to the observed peak magnitude in the rest-frame $B$ band, $\alpha$, $\beta$ and $M_B$ are nuisance parameters. According to \cite{C11}, $M_B$ may depend on the host stellar mass ($M_{\mathrm{stellar}}$) as
\begin{equation}
 M_B=\left\{\begin{array}{ll}
             M_B^1 & \mathrm{if}\,\,M_{\mathrm{stellar}}<10^{10}M_\odot.\\
             M_B^1+\Delta_M & \mathrm{otherwise.}\\
            \end{array}
\right.
\end{equation}

For SNe Ia from JLA, we have the likelihood $\like_{SN} \propto e^{-\frac{\chi^2_{SN}}{2}}$, where 
\begin{equation}
\chi^2_{SN} = \left[\hat{\bm{\mu}}(\theta_{SN})-{\bm\mu}(z,z_t,\theta_{mod,j})\right]^T\bm{C}^{-1}\left[\hat{\bm\mu}(\theta_{SN})-{\bm\mu}(z,z_t,\theta_{mod,j})\right]
\label{chi2SN}
\end{equation}
where $\theta_{SN}=(\alpha,\beta,M_B^1,\Delta_M)$, $\bm{C}$ is the covariance matrix of $\hat{\bm{\mu}}$ as described on \cite{JLA}, $\bm{\mu}(z,z_t,\theta_{mod,j})=5\log_{10}(d_L(z,z_t,\theta_{mod,j})/10\,\mathrm{pc})$ computed for a fiducial value $H_0=70$ km/s/Mpc.

\begin{figure}[ht]
 \includegraphics[width=\linewidth]{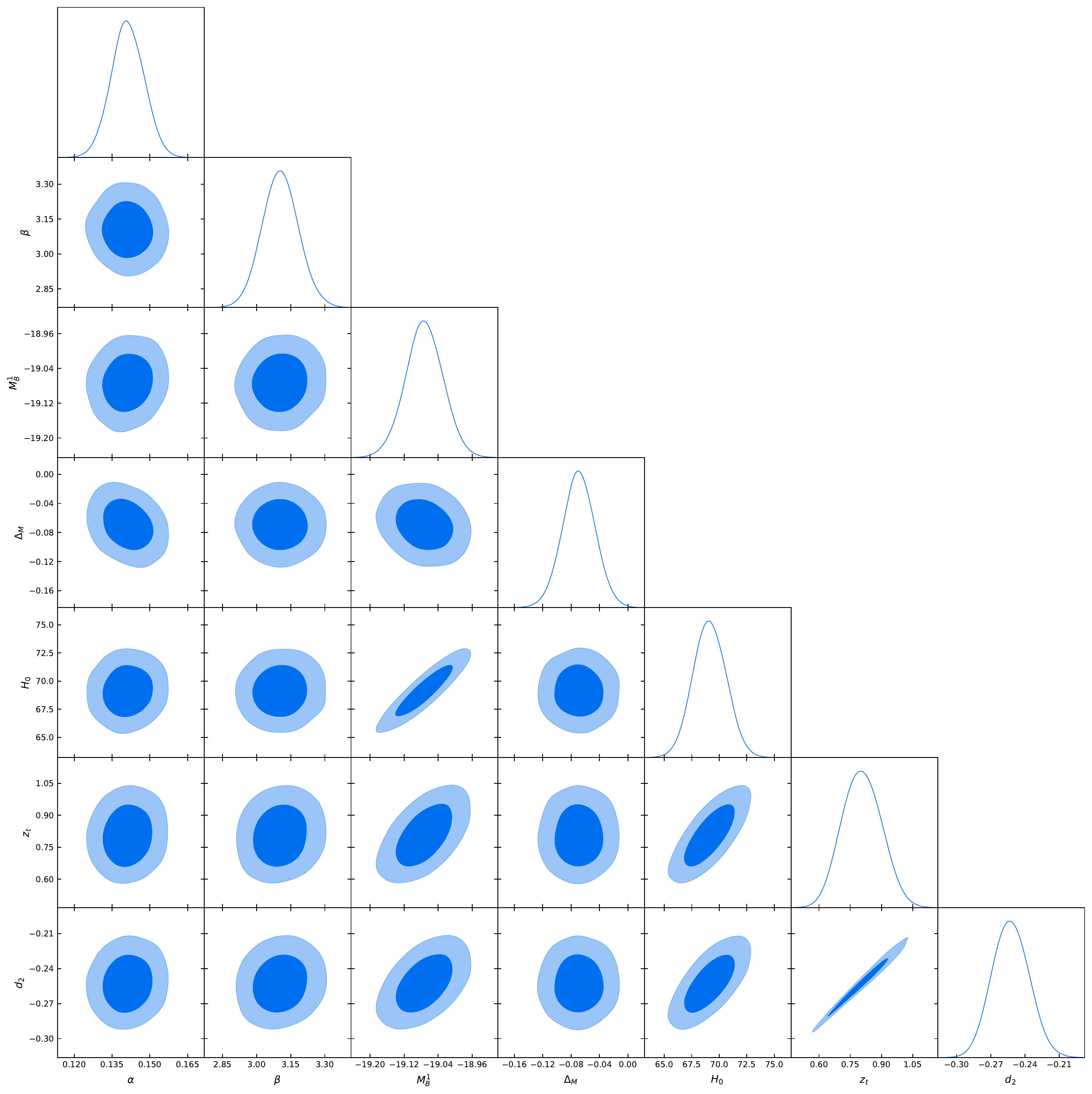}
 \caption{\label{DcPolinTriangAll} Combined constraints from JLA and $H(z)$ for $D_C(z)=z+d_2z^2+d_3z^3$.}
\end{figure}

\begin{figure}[ht]
 \includegraphics[width=\linewidth]{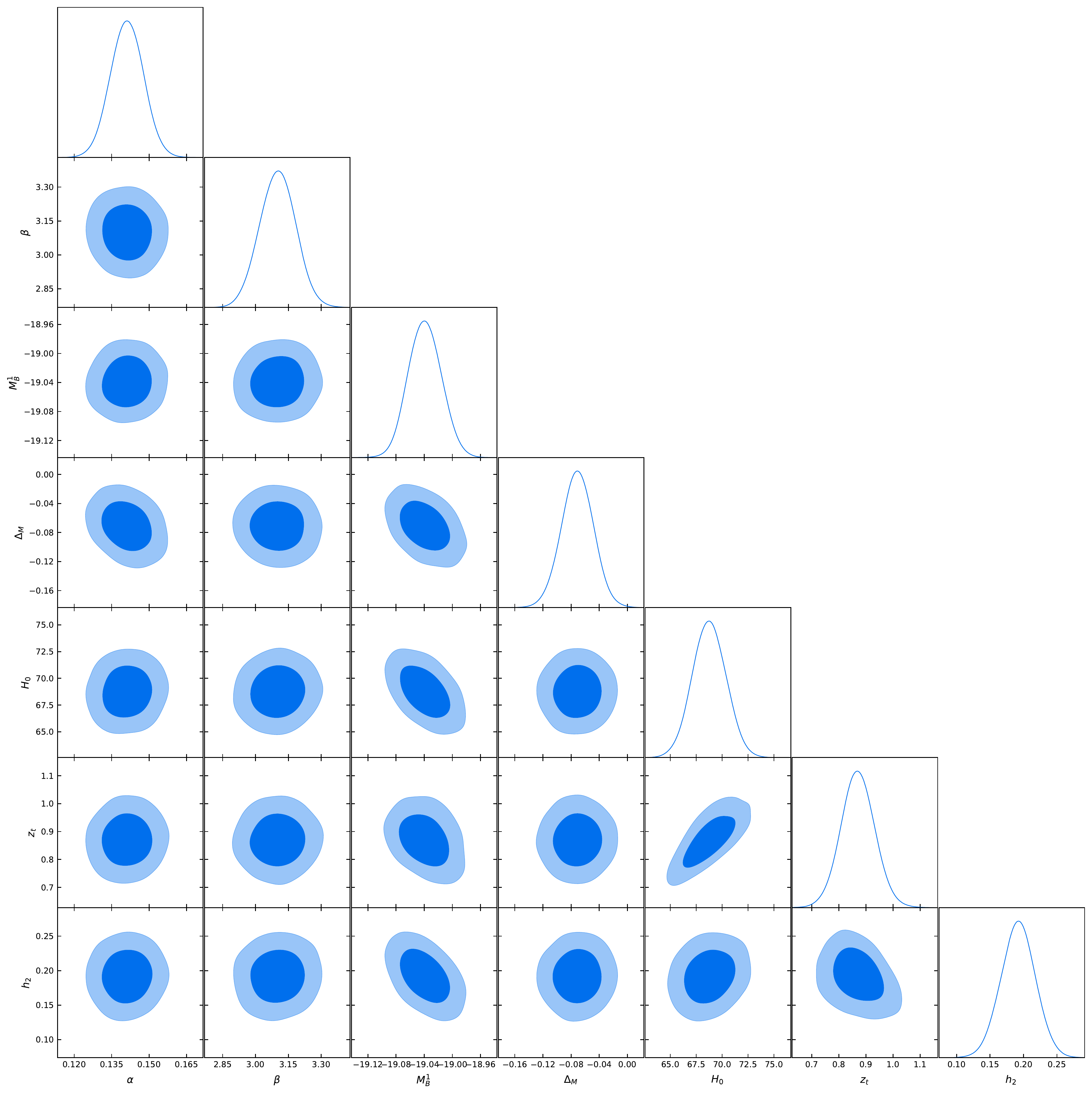}
 \caption{\label{HzPolinTriangAll} Combined constraints from JLA and $H(z)$ for $H(z)=H_0(1+h_1z+h_2z^2)$.}
\end{figure}

\begin{figure}[ht]
 \includegraphics[width=\linewidth]{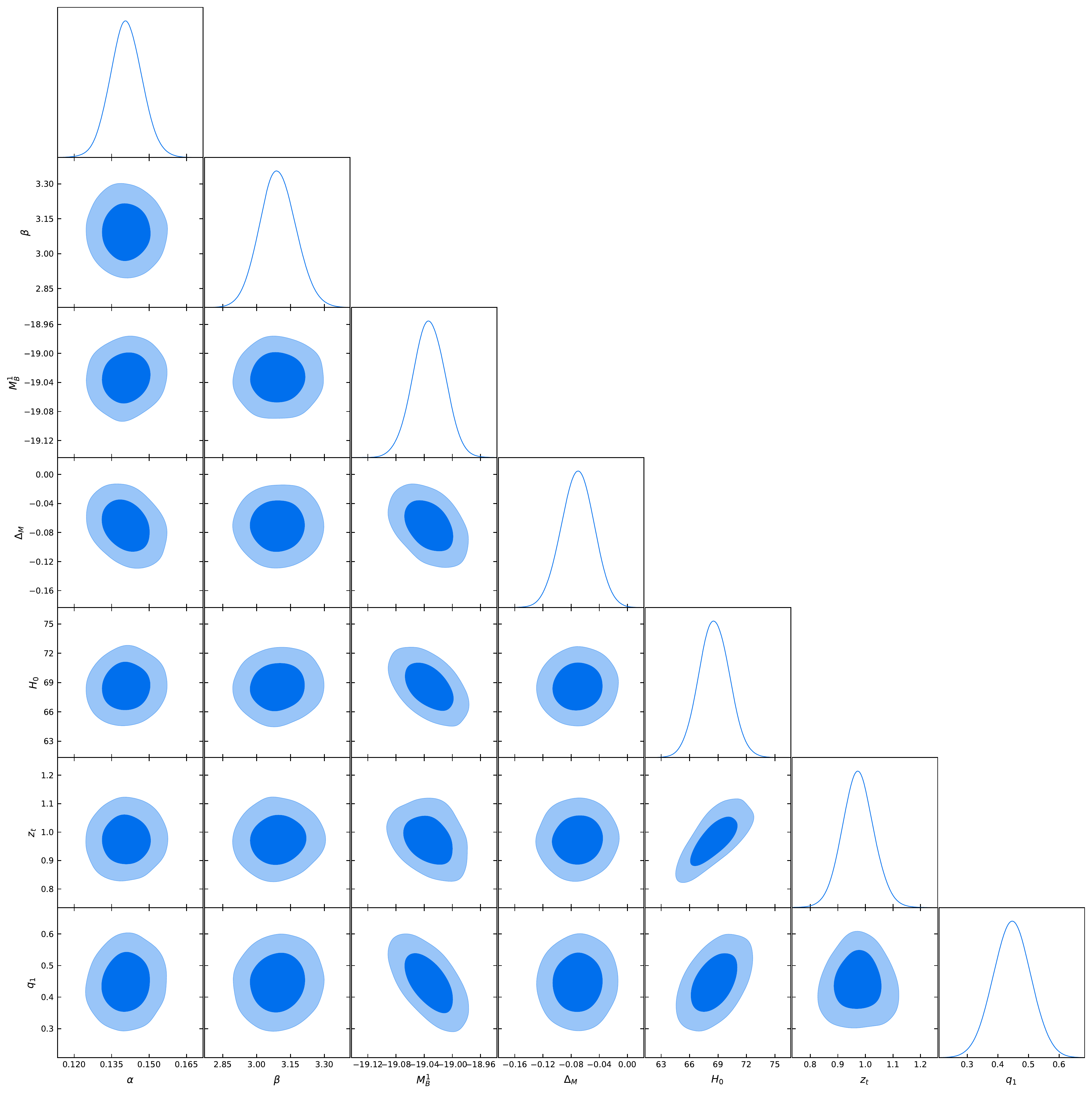}
 \caption{\label{q0q1TriangAll} Combined constraints from JLA and $H(z)$ for $q(z)=q_0+q_1z$.}
\end{figure}

In order to obtain the constraints over the free parameters, we have sampled the likelihood $\like\propto e^{-\chi^2/2}$ through Monte Carlo Markov Chain (MCMC) analysis. A simple and powerful MCMC method is the so-called Affine Invariant MCMC Ensemble Sampler by \cite{GoodWeare}, which was implemented in {\sffamily Python} language with the {\sffamily emcee} software by \cite{ForemanMackey13}. This MCMC method has advantage over the simple {Metropolis-Hastings} (MH) method, since it depends only on one scale parameter of the proposed distribution and also on the number of walkers, while MH method {is based} on the parameter covariance matrix, that is, it depends on $n(n+1)/2$ tuning parameters, where $n$ is the dimension of parameter space. {The main idea of the Goodman-Weare affine-invariant sampler is the so called ``stretch move'', where the position (parameter vector in parameter space) of a walker (chain) is determined by the position of the other walkers.} Foreman-Mackey {\it et al.} modified this method, in order to make it suitable for parallelization, by splitting the walkers in two groups, then the position of a walker in one group is determined {\it only} by the position of walkers of the other group\footnote{See \cite{AllisonDunkley13} for a comparison among various MCMC sampling techniques.}.

\begin{figure}[ht]
 \includegraphics[width=.8\linewidth]{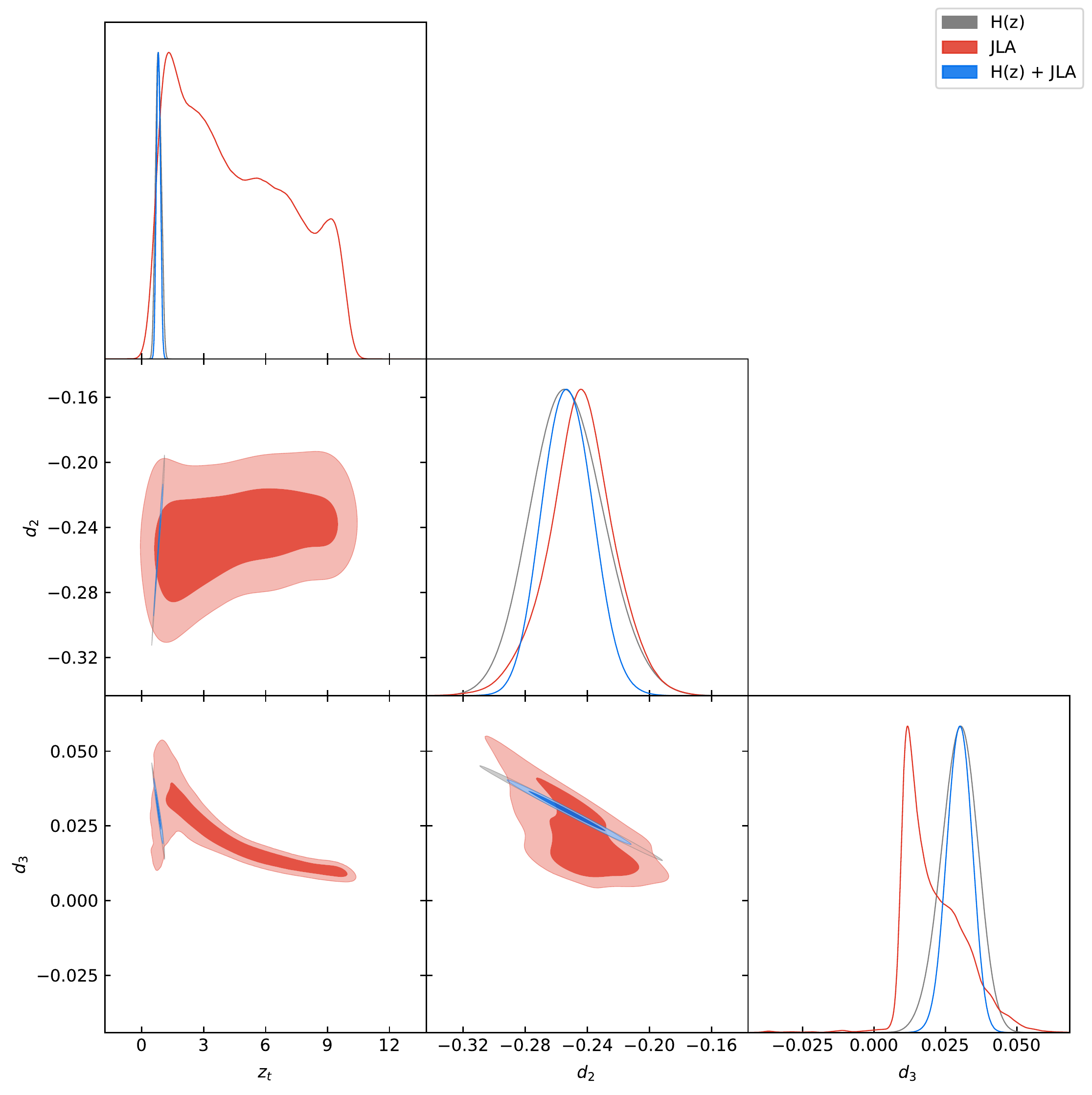}
 \caption{\label{DcPolinTriang}  {Constraints from JLA and $H(z)$ for $D_C(z)=z+d_2z^2+d_3z^3$.} }
\end{figure}

\begin{figure}[ht]
  \includegraphics[width=.8\linewidth]{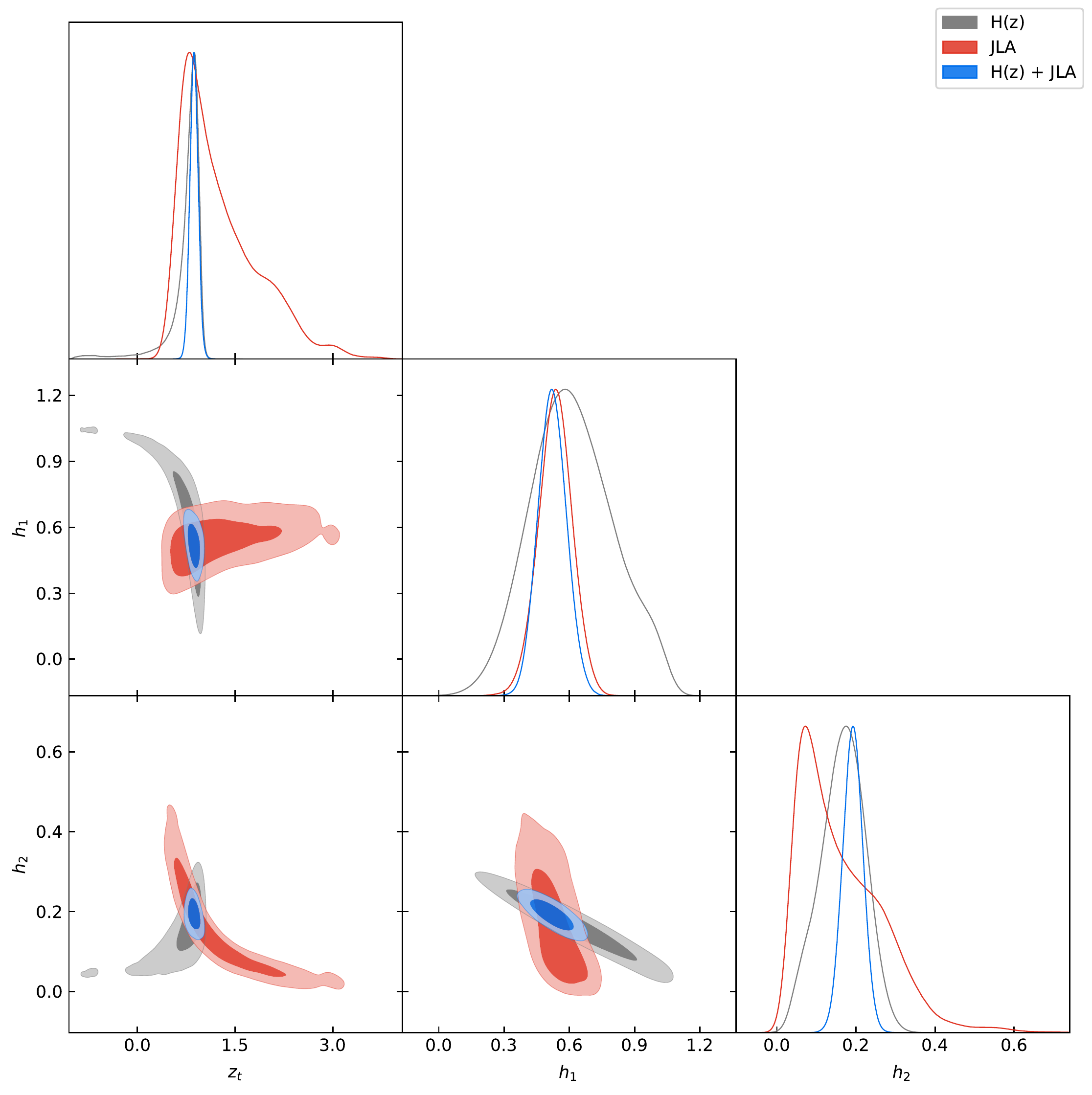}
  \caption{\label{HzPolinTriang} {Constraints from JLA and $H(z)$ for $H(z)=H_0(1+h_1z+h_2z^2)$.}}
\end{figure}



\begin{figure}[ht]
 \includegraphics[width=.8\linewidth]{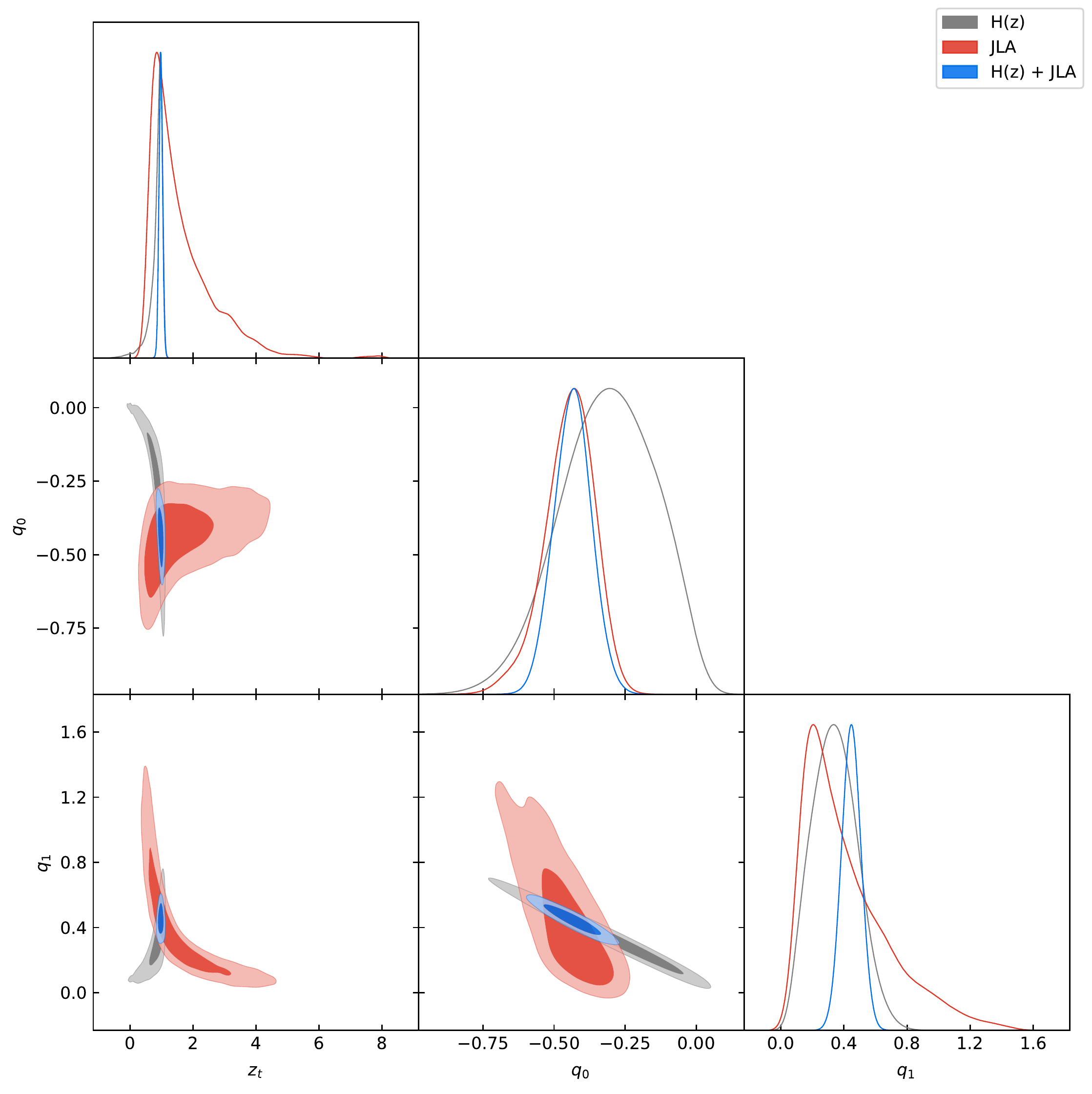}
 \caption{\label{q0q1Triang} {Constraints from JLA and $H(z)$ for $q(z)=q_0+q_1z$.}}
\end{figure}

\begin{figure}[ht]
\centering
 \includegraphics[width=.8\linewidth]{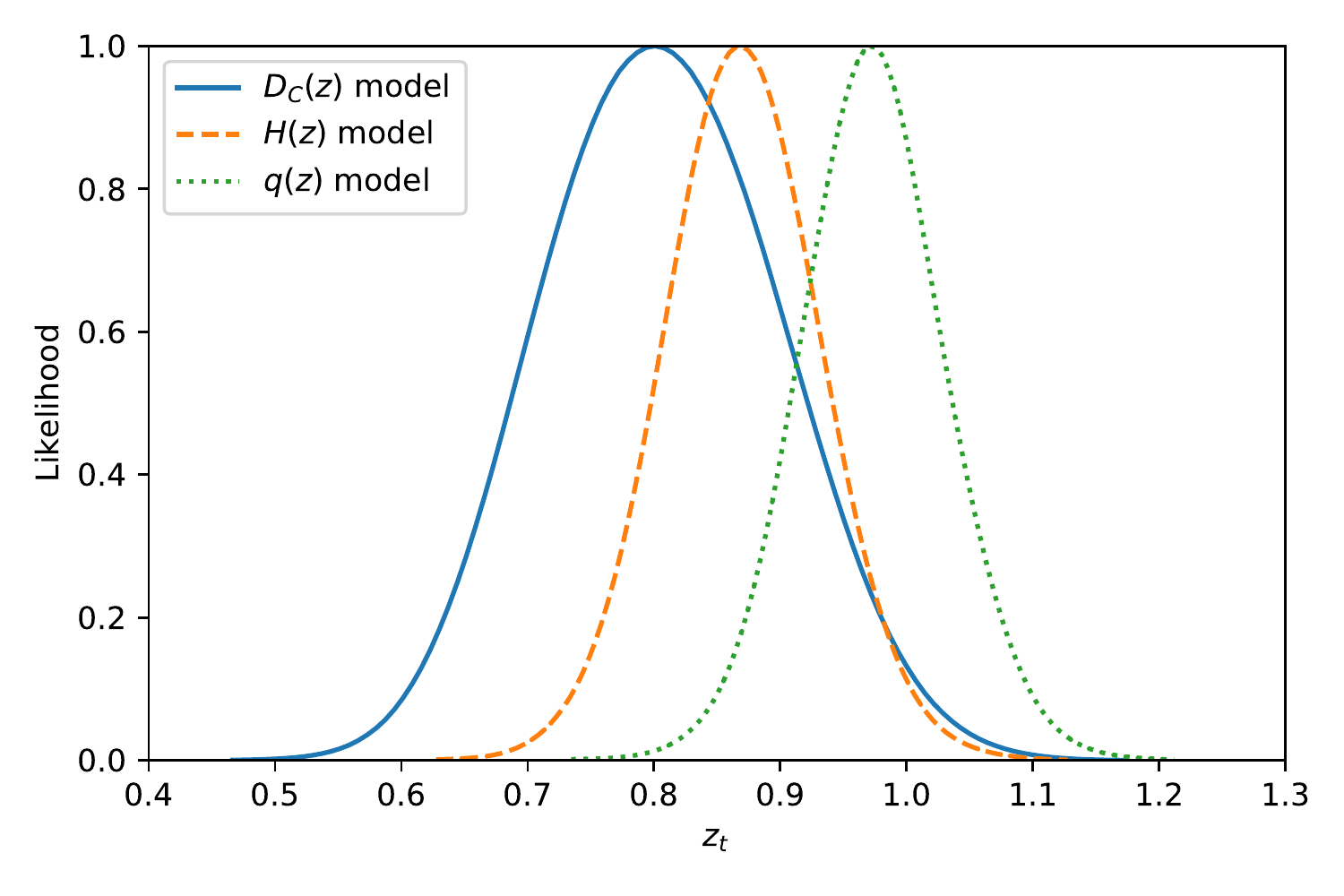}
 \caption{\label{ztlikes} Likelihoods for transition redshift from {JLA} and $H(z)$ data combined. Blue solid line corresponds to $D_C(z)$ parametrization, orange long-dashed line corresponds to $H(z)$ parametrization and green short-dashed line corresponds to $q(z)$ parametrization.}
\end{figure}

We used the freely available software {\sffamily emcee} to sample from our likelihood {in $n$-dimensional} parameter space. We have used flat priors over the parameters.
In order to plot all the constraints on each model in the same figure, we have used the freely available software {\sffamily getdist}\footnote{{\sffamily getdist} is part of the great MCMC sampler and CMB power spectrum solver {\sffamily COSMOMC}, by \cite{cosmomc}.}, in its {\sffamily Python} version. The results of our statistical analyses can be seen on Figs. \ref{DcPolinTriangAll}-\ref{ztlikes} and on Table \ref{tab1}.



In Figs.\ref{DcPolinTriangAll}-\ref{q0q1TriangAll}, we have the combined results for each parametrization. As one may see, we have always chosen $z_t$ as one of our fiducial parameter. The other parameters from each model is later obtained as derived parameters. As one may see in Figs.\ref{DcPolinTriangAll}-\ref{q0q1TriangAll}, the combination of JLA+$H(z)$ yields strong constraints over all parameters, especially $z_t$. Also, we find {negligible} difference in the SNe Ia parameters for each model. We have also to emphasize that the constraints over $H_0$ comes {just} from $H(z)$ {while} for JLA $H_0$ is fixed. We choose not to include other constraints over $H_0$ due to the recent tension from different limits over the Hubble constant. {At the end of this section, we compare our results with different constraints over $H_0$.}

In Figs.\ref{DcPolinTriang}-\ref{q0q1Triang}, we show explicitly the independent constraints from JLA and $H(z)$ over the cosmological parameters. As one may see in Fig.\ref{DcPolinTriang}, SNe Ia {sets weaker} constraints over $z_t$ for $D_C(z)$ parametrization. Almost all the $z_t$ constraint comes just from $H(z)$. For $d_2$, $H(z)$ and JLA yields similar constraints. For $d_3$, $H(z)$ yields {slightly} better constraints. For Figs.\ref{HzPolinTriang} and \ref{q0q1Triang}, the constraints over $z_t$ from SNe Ia are improved and one may see how SNe Ia and $H(z)$ complement each other in order to constrain the transition redshift. In Fig. \ref{HzPolinTriang}, one may see that the constraint over $h_1$ is better from JLA. The constraint over $h_2$ is better from $H(z)$. In Fig.\ref{q0q1Triang}, one may see that the constraint over $q_0$ is better from JLA. The constraint over $q_1$ is better from $H(z)$.

For all parametrizations, the best constraints over the transition redshift comes from $H(z)$ data, as first indicated by \cite{limajesus2014}. Moresco {\it et al.} \cite{moresco2016} also {found stringent} constraints over $z_t$ from $H(z)$ in their parametrization, however, they did not compare with SNe Ia constraints.


Fig.\ref{ztlikes} summarizes our combined constraints over $z_t$ for each parametrization. As one may see, the $q(z)$ model yields the strongest constraints over $z_t$. The other parametrizations are important for us to realize how much $z_t$ is still allowed to vary. All constraints are compatible at 1$\sigma$ c.l.

\newpage

Table \ref{tab1} shows the full numerical results from our statistical analysis. As one may see the SNe Ia constraints vary little for each parametrization. In fact,  we also have found constraints from the (faster) JLA binned data \cite{JLA}, however, when comparing with the (slower) full JLA constraints, we have found that the full JLA yields stronger constraints over the parameters, especially $z_t$. So we decided to deal only with the JLA full data.

\begin{table}[ht]
\begin{tabular}{|l|c|c|c|}
\hline
 Parameter                   &  $D_C(z)$                               & $H(z)$                                & $q(z)$                               \\
\hline
{\boldmath$\alpha         $} & $0.1412\pm 0.0065 \pm 0.013           $ & $0.1412\pm 0.0066   \pm 0.013       $ & $0.1408\pm 0.0064 \pm 0.013         $\\
{\boldmath$\beta          $} & $3.105\pm 0.080 \pm 0.16              $ & $3.101\pm 0.082 \pm 0.16            $ & $3.094\pm 0.080 \pm 0.16            $\\
{\boldmath$M_B^1          $} & $-19.073\pm 0.044^{+0.089}_{-0.090}   $ & $-19.039\pm 0.023^{+0.047}_{-0.045} $ & $-19.033\pm 0.023^{+0.045}_{-0.046} $\\
{\boldmath$\Delta_M       $} & $-0.069\pm 0.023\pm 0.046             $ & $-0.071\pm 0.023^{+0.045}_{-0.046}  $ & $-0.071\pm 0.023^{+0.046}_{-0.047}  $\\
{\boldmath$H_0            $} & $69.1\pm 1.5   \pm 3.0                $ & $68.8\pm 1.6 \pm 3.2                $ & $68.6\pm 1.6^{+3.3}_{-3.2}          $\\
{\boldmath$z_t            $} & $0.806\pm 0.094^{+0.19}_{-0.18}       $ & $0.870\pm 0.063^{+0.13}_{-0.12}     $ & $0.973\pm 0.058^{+0.12}_{-0.11}     $\\
{\boldmath$d_2            $} & $-0.253\pm 0.016^{+0.033}_{-0.031}    $ & --                                    & --                                   \\
$d_3                       $ & $0.0299\pm 0.0044^{+0.0085}_{-0.0090} $ & --                                    & --                                   \\
$h_1                       $ & --                                      & $0.522\pm 0.065 \pm 0.13            $ & --                                   \\
{\boldmath$h_2            $} & --                                      & $0.192\pm 0.026^{+0.051}_{-0.052}   $ & --                                   \\
$q_0                       $ & --                                      & --                                    & $-0.434\pm 0.065 \pm 0.13           $\\
{\boldmath$q_1            $} & --                                      & --                                    &$0.446\pm 0.062\pm 0.12              $\\
\hline
\end{tabular}
\caption{Constraints from JLA+$H(z)$ for $D_C(z)$, $H(z)$ and $q(z)$ parametrizations. The parameters without bold faces were treated as derived parameters. The central values correspond to the mean and the 1 $\sigma$ and 2 $\sigma$ c.l. correspond to the minimal 68.3\% and 95.4\% confidence intervals.}
\label{tab1}
\end{table}

By using 30 $H(z)$ data, plus $H_0$ from {Riess {\it et al}}. (2011) \cite{Riess11}, {Moresco {\it et al}}. \cite{moresco2016} found $z_t=0.64^{+0.1}_{-0.06}$ for $\Lambda$CDM and $z_t=0.4\pm0.1$ for their model independent approach. Only our $D_C(z)$ parametrization is compatible with their model-independent result. Their $\Lambda${CDM result} is compatible with all our parametrizations, although it is marginally compatible with $q(z)$.

{Another interesting result that can be seen in Table \ref{tab1} is the $H_0$ constraint. As one may see, the constraints over $H_0$ are consistent through the three different parametrizations, with a little smaller uncertainty for $D_C(z)$, $H_0=69.1\pm1.5$ km/s/Mpc.} The constraints over $H_0$ are quite stringent today from many observations \cite{RiessEtAl16,Planck16}. However, there is some tension among $H_0$ values estimated from Cepheids \cite{RiessEtAl16} and from CMB \cite{Planck16}. While Riess {\it et al.} advocate $H_0=73.24\pm1.74$ km/s/Mpc, the Planck collaboration analysis yields $H_0=66.93\pm0.62$ km/s/Mpc, a 3.4$\sigma$ lower value. It is interesting to note, from our Table \ref{tab1} that, although we are working with model independent parametrizations and data at intermediate redshifts, our result is in better agreement with the high redshift result from Planck. In fact, all our results are compatible within 1$\sigma$ with the Planck's result, while it is incompatible at 3$\sigma$ with the Riess' result.



\section{\label{conclusion}Conclusion}

{The accelerated expansion of Universe is confirmed by different sets of cosmological observations. Several models proposed in literature satisfactorily explain the transition from decelerated phase to the current accelerated phase. A more significant question is when the transition occurs from one phase to another,} and the parameter that measures this transition is called the transition redshift, $ z_t $. The determination of $ z_t $ is strongly dependent on the cosmological model adopted, thus the search for methods that allow the determination of such parameter in a model independent way are of fundamental importance, since it would serve as a test for several cosmological models.

In the present work, we wrote the comoving distance $ D_C $,  the Hubble parameter $H (z) $ and  the deceleration parameter $q(z)$ as  third, second and first degree polynomials on $z$, respectively {(see equations (\ref{DcPolin}), (\ref{EzPolin}) and (\ref{eqqz}))}, and obtained, for each case, the $ z_t $ value. Only a flat universe was assumed and the estimates for $z_t$ were obtained, independent  of a specific cosmological model.  As observational {data,} we have used  Supernovae type Ia and Hubble parameter measurements. Our results can be found in {Figures \ref{DcPolinTriangAll}-\ref{q0q1Triang}. As one may see from Figs. \ref{DcPolinTriang}-\ref{q0q1Triang}}, the analyses by using SNe Ia (red color) and $H(z)$ data (blue color) are complementary to each other, providing tight limits in the parameter spaces. {As a result, the values obtained for the transition redshift in each case were $0.806\pm 0.094$, $0.870\pm 0.063$ and $0.973\pm 0.058$ at 1$\sigma$ c.l., respectively (see Fig. \ref{ztlikes}).}

\begin{acknowledgments}
JFJ is supported by Funda\c{c}\~ao de Amparo \`a Pesquisa do Estado de S\~ao Paulo - FAPESP ({Process no. 2017/05859-0}). RFLH acknowledges financial support from  {Conselho Nacional de Desenvolvimento Cient\'ifico e Tecnol\'ogico} (CNPq) and UEPB (No. 478524/2013-7, 303734/2014-0). SHP is grateful to CNPq, for financial support (No. 304297/2015-1, 400924/2016-1).
\end{acknowledgments}


\end{document}